\documentclass[twocolumn,pre,showpacs,amsmath,amssymb,floatfix]{revtex4-1}

\usepackage{graphicx}
\usepackage{dcolumn}
\usepackage{bm}
\usepackage[utf8]{inputenc}
\usepackage{amsthm}
\usepackage{mathrsfs}
\usepackage{framed}
\usepackage[dvipsnames]{xcolor}\usepackage{empheq}
\usepackage[colorlinks=true,citecolor=blue,linkcolor=blue,urlcolor=blue]{hyperref}

\newtheorem{theorem}{Theorem}

\newcommand{\ens}[1]{\langle #1 \rangle}

\definecolor{light_blue}{rgb}{0.8, 0.888151, 1.}
\colorlet{shadecolor}{light_blue}
\newcommand*\mybluebox[1]{ 
\colorbox{shadecolor}{\hspace{1em}#1\hspace{1em}}}   

\DeclareMathAlphabet{\mathdutchcal}{U}{dutchcal}{m}{n}
\SetMathAlphabet{\mathdutchcal}{bold}{U}{dutchcal}{b}{n}
\DeclareMathAlphabet{\mathdutchbcal}{U}{dutchcal}{b}{n}

\begin{document}

\preprint{}

\title{Statistical Mechanics of Discrete Multicomponent Fragmentation}

\author{Themis Matsoukas}
\email{txm11@psu.edu}
\affiliation{Department of Chemical Engineering, The Pennsylvania State University, University Park, PA 16802.}

\date{\today}
\begin{abstract}
We formulate the statistics of the discrete multicomponent fragmentation event using a methodology borrowed from statistical mechanics. We generate the ensemble of all feasible distributions that can be formed when a single integer multicomponent mass is broken into fixed number of fragments and calculate the combinatorial multiplicity of all distributions in the set. We define random fragmentation by the condition that the probability of distribution be proportional to its multiplicity and obtain the partition function and the mean distribution in closed form. We then introduce a functional that biases the probability of distribution to produce in a systematic manner fragment distributions that deviate to any arbitrary degree from the random case. We corroborate the results of the theory by Monte Carlo simulation and demonstrate examples in which components in sieve cuts of the fragment distribution undergo preferential mixing or segregation relative to the parent particle. 
\end{abstract}
\keywords{discrete fragmentation, multicomponent, partition function, multiplicity of distribution}
\maketitle
\section{Introduction}
Objects disintegrate into fragments via impact, detonation, degradation of structure or cleavage of the bonds that hold the structure together. Objects may range from molecules and cells to macroscopic masses to celestial bodies and the physical mechanisms by which fragmentation occurs are as diverse. This diversity of scale and physics is united by a common mathematical description based on the size distribution of the fragments that are produced. 
Fragmentation has been studied extensively in particular with respect to modeling the evolution of the size distribution \citep{Ramkrishna:00}, a problem that commonly arises in polymeric, colloidal and granular materials as well as in biological populations where fragmentation refers to the splitting of groups of animals, for example \citep{Gueron:MB95}. It is a rich mathematical subject that in addition to its practical relevance to scientific and engineering problems exhibits under certain conditions a remarkable transition, shattering, a process akin to a phase transition \citep{Connaughton:PR18,PhysRevE.68.021102,Ernst:JPMG93,Ziff:JPMG92}. 
Several analytic results have been given in the literature in the context of population balances under various models in the discrete and continuous domain \citep{Montroll:JCP40,Simha:JAP41,Simha:JCP56,Ziff:JPMG85,Ziff:M86,McGrady:PRL87,Ziff:JPMG91}. In these studies fragmentation is nearly always binary  and the particle consists of a single component. In one-component fragmentation the distribution of fragments is a function of a single variable, ``size,'' which usually refers to the mass of the fragment. A multidimensional case was considered by \citet{Krapivsky:PR94} who considered the fragmentation of objects in $d$-dimensional space. A treatment of multi-nary fragmentation was given by \citet{Simha:JAP41,Simha:JCP56} who studied fragmentation in the context of polymer degradation. With the exception of Simha who used combinatorial arguments to obtain the distribution of fragments in linear \citep{Montroll:JCP40} and branched \citep{Simha:JCP56} chains, most studies adopt either binary random aggregation, which produces a uniform distribution of fragments, or some empirical distribution of fragments that allows arbitrary number of fragments and non-random distribution of sizes (see for example \citet{Austin:IECPDD76,Ziff:JPMG91}). 

Multicomponent fragmentation with no limitation on the number of fragments is the rule rather than the exception in most problems of practical interest. This cannot be obtained by simple extension of the one-component problem. In addition to the size distribution of the fragments we must also tackle the distribution of components and provide rules (a ``model'') for apportioning components to the fragments. We must offer a definition of what is meant by ``random fragmentation'' when both size and composition are distributed  and provide the means for constructing models that deviate from the random case to any  desired extent. 

The purpose of this paper is to formulate the statistics of a single fragmentation event in the discrete domain for arbitrary number of fragments and components and to provide the means of connecting this mathematical formalism to physics. The main idea is this: We start with a multicomponent particle that contains discrete units of multiple components, subject it to one fragmentation event into fixed number of fragments and construct the set of all fragment distributions that can be obtained. We calculate the partition function of this ensemble of random fragments, assign probabilities in proportion to the multiplicity of each distribution and obtain the mean distribution in terms of the partition function. We then introduce a bias functional that biases the distribution away from that of random fragmentation. We present results from  Monte Carlo simulations to corroborate the theory and show that components may preferentially mix or unmix in the fragments depending on the choice of the bias functional. 
\clearpage

\section{Random fragmentation}
\subsection{One-component random fragmentation}

In discrete fragmentation, a particle composed of $m$ integer units breaks up into $N$ fragments, $\{f_1,f_2\cdots f_N\}$ that satisfy the mass balance conditions
\begin{equation}
   \sum_{i=1}^{N} f_i = m . 
\end{equation}
We construct the distribution of fragments $\mathbf{n}=(n_1,n_2\cdots)$, such that $n_i$ is the number of fragments that contain $i$ units of mass. We suppose that $N$ is fixed but $\mathbf{n}$ is not; that is, if the fragmentation event is repeated with an identical parent particle the distribution of fragments may be different but the total number of fragments is always $N$. We refer to this process as $N$-nary fragmentation.  All fragment distributions produced by this mechanism satisfy the following two conditions:
\begin{align}
\label{constraint_1}
   \sum_{i=1}^\infty n_i = N, \\
\label{constraint_2}
   \sum_{i=1}^\infty i n_i = m .
\end{align}
The first condition states that the number of fragments is $N$; the second that their  mass is equal to the mass of the parent particle. Conversely, any distribution that satisfies the above two equations is a feasible distribution of fragments by $N$-nary fragmentation of mass $m$. Thus the set $\mathscr E_{m,N}$ of all distributions that satisfy Eqs.\ (\ref{constraint_1}) and (\ref{constraint_1}) forms the ensemble of fragment distributions produced from $m$. 

We will call the process \textit{random fragmentation} if all ordered lists of $N$ fragments produced by the same mass are equally probable.  This views the ordered list of fragments, which we call configuration, as the primitive stochastic variable in this problem. 

\subsubsection{Probability of random fragment distribution}

\begin{figure}
\begin{center}
\includegraphics[width=3.25in]{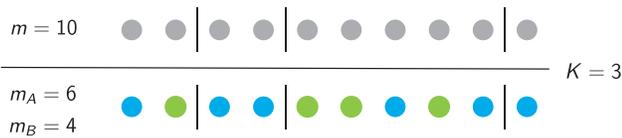}
\end{center}
\caption{Random fragmentation of integer mass $m$ into $N$ pieces is equivalent to breaking a string with $m$ beads at $N-1$ random points. If the mass is made up of two colors (components) every permutation of the beads is equally probable. }
\label{fig1}
\end{figure}

\begin{theorem}
The probability of distribution $\mathbf{n}$ produced by random $N$-nary fragmentation of mass $m$ is
\begin{equation}
\label{prob_dstr}
   P(\mathbf{n}) = \frac{\mathbf{n}! }{\displaystyle\binom{m-1}{N-1}}, 
\end{equation}
where $\mathbf{n}! $ is the multinomial coefficient of $\mathbf{n} =(n_1,n_2\cdots)$
\begin{equation}
   \mathbf{n}!  = \frac{(\sum_i n_i)!}{\prod_i n_i!}
   = \frac{N!}{n_1! n_2!\cdots} . 
\end{equation}
\proof First we note that the number of ordered lists (in all permutations) that can be formed by breaking integer $m$ into $N$ fragments is 
\begin{empheq}[box=\mybluebox]{equation}
\label{partition_function}
   \Omega^{(1)}_{m;N} = \binom{m-1}{N-1} .
\end{empheq}
This is the number of ways to partition integer $m$ into $N$ parts and can be shown easily as follows \citep{Bona:2006}: thread $m$ balls into a string and partition them into $N$ pieces by cutting the string at $N-1$ points (Fig.\ \ref{fig1}). There are $m-1$ points where we can cut and must choose $N-1$ of them. The number of ways to do this is the binomial factor on the RHS of  Eq.\ (\ref{partition_function}). 

If all ordered lists of fragments are equally probable, the probability of ordered list $\mathbf{m} = (m_1,m_2\cdots m_N)$ is
\begin{equation}
   P(\mathbf{m}) = \frac{1}{\Omega^{(1)}_{m;N}} . 
\end{equation}
There are $\mathbf{n}! $ ordered lists with the same distribution of fragments $\mathbf{n}$.  Accordingly, the probability of $\mathbf{n}$ is
\begin{equation}
   P(\mathbf{n}) = \mathbf{n}! P(\mathbf{m}) = \frac{\mathbf{n}! }{\Omega^{(1)}_{m;N}}. 
\end{equation}
where $\mathbf{n}$ is the fragment distribution in $\mathbf{m}$.  This proves the theorem. \qed 
\end{theorem}

The multinomial factor $\mathbf{n!}$ represents the multiplicity of distribution $\mathbf{n}$, namely, the number of configurations (ordered lists of fragments) represented by $\mathbf{n}$. Using $\omega(\mathbf{n}) = \mathbf{n!}$ to notate the multiplicity of distribution, the probability of distribution is expressed as
\begin{equation}
   P(\mathbf{n}) = \frac{\omega(\mathbf{n})}{\Omega^{(1)}_{m;N}},
\end{equation}
and $\Omega^{(1)}$ satisfies
\begin{equation}
\label{sum_of_multi}
   \sum_\mathbf{n} \omega(\mathbf{n}) = \Omega^{(1)}_{m;N} .  
\end{equation}
The summation over all distributions $\mathbf{n\in\mathscr E_{m,N}}$, namely, all distributions that satisfy Eqs.\ (\ref{constraint_1}) and (\ref{constraint_2}). Accordingly, $\Omega^{(1)}_{m;N}$ is the total multiplicity in the ensemble, equal to the number of ordered configurations of fragments that can be produced from integer mass $m$ breaking into $N$ fragments. We refer to $\Omega^{(1)}_{m;N}$ as the partition function of the one-component ensemble of fragments.

\subsubsection{Mean fragment distribution}

Each distribution $\mathbf{n}$ appears in the ensemble of fragment distributions  with probability $P(\mathbf{n})$; the mean distribution of fragments is their ensemble average:
\begin{equation}
   \ens{\mathbf{n}} = \sum_{\mathbf{n}} \mathbf{n}\, P(\mathbf{n})
\end{equation}
with $P(\mathbf{n})$ from Eq.\ (\ref{prob_dstr}) and with the summation going over all distributions that are produced by $N$-nary fragmentation of integer mass $m$. 

\begin{theorem}
The mean distribution in $N$-nary random fragmentation is 
\begin{empheq}[box=\mybluebox]{equation}
\label{mean_dstr}
   \frac{\ens{n_k}}{N} = \left.\binom{m-k-1}{N-1}\right/\binom{m-1}{N-1} . 
\end{empheq}
with $k=1,\cdots m-N+1$. 
\proof First we write the probability of distribution in the form
\begin{equation}
\label{prob_b2}
   P(\mathbf{n}) = \frac{N!}{\Omega^{(1)}_{m;N}}
   \prod_{i=1}^\infty \frac{\alpha_i^{n_i}}{n_i!} 
\end{equation}
with $\alpha_i>0$ and note that this reverts to Eq.\ (\ref{prob_dstr}) when $\alpha_i=1$. We will retain the factors $\alpha_i$ and will set them equal to 1 at the end. The normalization condition on the probability $P(\mathbf{n})$ reads
\begin{equation}
\label{partition_function2}
   \Omega^{(1)}_{m;N} = N!\sum_{\mathbf{n}} \prod_{i=1}^\infty \frac{\alpha_i^{n_i}}{n_i!}  .
\end{equation}
The derivative of $\log \Omega^{(1)}_{m;N}$ with respect to $n_k$ is
\begin{equation}
\label{dlog_omega_dak_1}
   \frac{d\log \Omega^{(1)}_{m;N}}{d \alpha_k}
   = 
   \frac{\alpha_k N!}{\Omega^{(1)}_{m;N}} \sum_{\mathbf{n}} n_k  
   \prod_i\left(\frac{\alpha_i^{n_i}}{n_i^!}\right)
   =
   \alpha_k \ens{n_k} ,
\end{equation}
where $\ens{n_k}$ is the mean value of $n_k$ in the ensemble of fragments.  We also have
\begin{multline}
\label{dlog_omega_dak_2}
   \frac{d\Omega^{(1)}_{m;N}}{d\alpha_k}
   =
   N \left\{(N-1)! \sum_{\mathbf{n}}\left(\cdots 
   \frac{\alpha_i^{n_k-1}}{(n_k-1)!}\cdots\right)\right\}
   \\
   =
   N \Omega^{(1)}_{m-k;N-1} .
\end{multline}
The summand in the expression in the middle amounts to removing one fragment of mass $k$ from all distributions of the ensemble; accordingly, the quantity in braces is the partition function $\Omega^{(1)}_{m-k;N-1}$. Combining Eqs.\ (\ref{dlog_omega_dak_1}) and (\ref{dlog_omega_dak_2}) and setting $\alpha_k=1$  we obtain Eq.\ (\ref{mean_dstr}). \qed
\end{theorem}
Equation (\ref{mean_dstr}) was previously obtained by \citet{Montroll:JCP40} via a  combinatorial derivation. Notably it is the same distribution as in discrete binary aggregation (the \textit{reverse} process of binary fragmentation) with constant kernel, derived by \citet{Hendriks:ZPCM85B} who also credit older unpublished work by White. 
%
%
For large $m$ the fragment distribution becomes
\begin{equation}
   n_k \to \frac{N(N-1)}{m}\left(1-\frac{k}{m}\right)^{N-2} .
\end{equation}
This is the continuous limit of random fragmentation of a straight line into $N$ segments and is an elementary result that has been derived multiple times in the literature. The earliest report known to us is by \citet{Feller:PR40} who corrected an earlier approximation by \citet{Ruark:PR39}. 


\subsection{Two-component random fragmentation}
\subsubsection{Representations of bicomponent populations}
\paragraph{The bicomponent distribution}
We now consider a particle that is made of two components. The particle contains $m_A$ units of component $A$, $m_B$ units of component $B$ and its mass is $m=m_A+m_B$.  The distribution of fragments is given by the two dimensional vector $\mathbf{n} = \{n_{a,b}\}$ where $n_{a,b}$ is the number of fragments that contain $a$ units of $A$ and $b$ units of $B$. This distribution satisfies the conditions
\begin{align}
\label{bi_constraint1}
   & \sum_{a=0}^\infty\sum_{b=0}^\infty n_{a,b} = N, \\
\label{bi_constraint2}
   & \sum_{a=0}^\infty\sum_{b=0}^\infty a n_{a,b} = m_A, \\
\label{bi_constraint3}
   & \sum_{a=0}^\infty\sum_{b=0}^\infty b n_{a,b} = m_B .
\end{align}
The set $\mathscr E_{m_A, m_B;N}$ of all distributions that satisfy the above conditions constitutes the set of feasible distributions in bicomponent fragmentation.

\paragraph{Color-blind distribution}

The color-blind size distribution or simply ``size distribution'' $\mathbf{n}_{A+B} = \{n_k\}$ is the distribution of the mass of the fragments $k=a_b$ regardless of composition:
\begin{equation}
   n_k = \sum_{a=0}^{k} n_{a,k-a},\quad k=1,2\cdots
\end{equation}
and satisfies the conditions
\begin{align}
   & \sum_{k=1}^\infty n_i = N, \\
   & \sum_{k=1}^\infty i n_i = m_A+m_B = m. 
\end{align}
These are the same as Eqs.\ (\ref{constraint_1}) and (\ref{constraint_2}) in the one-component case for a particle with mass $m_A+m_B$. Accordingly, the feasible set of the color-blind distribution is $\mathscr E_{m_A+m_B; N}$. 

\paragraph{Sieve-cut distribution}

The sieve-cut distribution $\mathbf{n}_{A|k} = \{n_{a|k}\}$ is the number of fragments with size $k$ that contains $a$ units of component $A$:
\begin{equation}
   n_{a|k} = n_{a,k-a},\quad
   (a=1\cdots k,~k=1\cdots\infty) .
\end{equation}
and satisfies the normalizations
\begin{align*}
   & \sum_{a=0}^k\sum_{k=1}^\infty n_{a|k} = N , \\
   & \sum_{a=0}^k\sum_{k=1}^\infty a n_{a|k} = m_A , \\
   & \sum_{a=0}^k\sum_{k=1}^\infty k n_{a|k} = m_A+m_B .
\end{align*}
Normalizing the sieve-cut distribution by the number of fragments of size $k$ we obtain the compositional distribution of component $A$ within fragments of fixed size $k$, 
\begin{equation}
   c_{a|k} = \frac{n_{a|k}}{n_k} . 
\end{equation}
The compositional distribution is normalized to unity and may be interpreted as a conditional probability: it is the probability to obtain a fragment with $a$ units of $A$ given that the fragment has mass $k$. 
The bicomponent distribution may now be expressed in terms of the color-blind distribution $\mathbf{n}_{A+B}$ and the compositional distribution $c_{a|k}$ in the form
\begin{equation}
\label{dstr_conditional}
   n_{a,k-a} = n_k c_{a|k} .
\end{equation}
If we divide both sides by the total number of fragments the result reads as a joint probability: The probability $n_{a,k-a}/N$ to obtain a fragment with mass $k$ that contains $a$ units of component $A$ is equal to the probability $n_k/N$ to obtain a fragment of mass $k$ times the probability $c_{a|k}$ to obtain a fragment with $a$ units of component $A$ given that the mass of the fragment is $k$. 

\subsubsection{The ensemble of random fragment distributions}

Random fragmentation in implemented by analogy to the one-component case: we line up the unit masses in the particle into a string and cut at $N-1$ places. Every cut is equally probable and so is every permutation in the order of the beads. 

\begin{theorem}
The probability of fragment distribution $\mathbf{n}$ is
\begin{empheq}[box=\mybluebox]{equation}
\label{prob_dstr_bi}
   P(\mathbf{n}) = \frac{\mathbf{n!}}{\Omega^{(2)}_{m_A,m_B;N}}
   \prod_{a=0}^{\infty}
   \prod_{b=0}^{\infty}
   \binom{a+b}{a}^{n_{a,b}} .
\end{empheq}
where $\mathbf{n!}$ is the multinomial coefficient of the bicomponent distribution,
\begin{equation}
\label{multinomial_muti}
   \mathbf{n!} = \frac{N!}{\prod_{a=0}^\infty\prod_{b=0}^\infty n_{a,b}!} 
\end{equation}
and $\Omega^{(2)}_{m_A,m_B;N}$ is the two-component partition function, given by
\begin{empheq}[box=\mybluebox]{equation}
\label{partition_function_bi}
   \Omega^{(2)}_{m_A,m_B;N} = \binom{m_A+m_B}{m_A}\Omega^{(1)}_{m_A+m_B;N} .
\end{empheq}
\proof
First we count the number of ordered sequences of fragments (configurations). Configurations are distinguished by the order the fragments and by the order of  components within fragments (Fig.\ \ref{fig1}). We color the components and place them in a line in some order. There are $m_A$ $A$'s and $m_B$ $B$'s; the number of permutations is $\binom{m_A+m_B}{m_A}$. Each permutation produces $\Omega^{(1)}_{m_A+m_B;N}$ configurations with $\Omega^{(1)}$ given in Eq.\ (\ref{partition_function}).  The total number of configurations therefore is their product and proves  Eq.\ (\ref{partition_function_bi}).  

Since all configurations are equally probable the probability of fragment distribution $\mathbf{n}$ is proportional to the number of configurations with that distribution. This is equal to the number of permutations in the order of the fragments and in the order of components within fragments. The number of permutations in the order of fragments is given by the multinomial factor of bicomponent distribution in Eq.\ (\ref{multinomial_muti}). The number of permutations of components within a fragment that contains $a$ units of $A$ and $b$ units of $B$ is $\binom{a+b}{a}$ and since there are $n_{a,b}$ such fragments, the total number of internal permutations in distribution $\mathbf{n}$ is
\begin{equation}
\label{multi_int}
   \prod_{a=0}^\infty\prod_{b=0}^\infty\binom{a+b}{a}^{n_{a,b}} .
\end{equation}
The probability of distribution $\mathbf{n}$ is equal to the  product of Eqs (\ref{multinomial_muti}) and (\ref{multi_int}) divided by the total number of configurations, given by Eq.\ (\ref{partition_function_bi}):
\begin{equation}
\label{prob_dstr_bi1}
   P(\mathbf{n}) = \frac{\mathbf{n!}}{\Omega^{(2)}_{m_A,m_B;N}}
   \prod_{a=0}^{\infty}
   \prod_{b=0}^{\infty}
   \binom{a+b}{a}^{n_{a,b}} .
\end{equation}
This proves the theorem.\qed
\end{theorem}
%
As a corollary we obtain the multiplicity of the bicomponent distribution,
\begin{equation}
   \omega(\mathbf{n}) = \mathbf{n!}    
   \prod_{a=0}^{\infty}
   \prod_{b=0}^{\infty}
   \binom{a+b}{a}^{n_{a,b}} .
\end{equation}
Thus we write
\begin{equation}
   P(\mathbf{n}) 
   =\frac{\omega(\mathbf{n})}{\Omega^{(2)}_{m_A,m_B;N}}
\end{equation}
with $\Omega^{(2)} = \sum_\mathbf{n} \omega(\mathbf{n})$. 

An alternative equation for $P(\mathbf{n})$ is obtained by expressing the bicomponent distribution $\mathbf{n}$ in terms of the color-blind distribution $\mathbf{n}_{A+B}$ and all sieve-cut distributions $\mathbf{n}_{A|k}$. The result is 
\begin{empheq}[box=\mybluebox]{equation}
\label{prob_dstr_bi2}
   P(\mathbf{n}) = 
   \frac{\mathbf{n}_{A+B}!}{\Omega^{(2)}_{m_A,m_B;N}}
   \prod_{k=0}^\infty\left\{
   \mathbf{n}_{A|k}! 
   \prod_{a=0}^k \binom{k}{a}^{n_{a|k}}
   \right\}    
\end{empheq}
and is based on the identity 
\begin{equation}
\label{multiplicity_identity}
   \mathbf{n!}\prod_{a=0}^\infty\prod_{b=0}^\infty\binom{k}{a}^{n_{a,b}} 
   = \mathbf{n}_{A+B}!
     \prod_{k=0}^\infty\left\{
     \mathbf{n}_{A|k}! 
     \prod_{a=0}^k \binom{k}{a}^{n_{a|k}}
     \right\}    
\end{equation}
where $\mathbf{n}_{A+B}!$ is the multinomial coefficient of the color-blind distribution,
\begin{equation}
   \mathbf{n}_{A+B}! = \frac{N!}{n_1! n_2! \cdots} 
\end{equation}
and $\mathbf{n}_{A|k}!$ is the multinomial coefficient of the sieve-cut distribution,
\begin{equation}
   \mathbf{n}_{A|k}! = \frac{n_k!}{n_{0|k}! n_{1|k}!\cdots n_{k|k}!} . 
\end{equation}
%

\subsubsection{Mean fragment distribution}
\begin{theorem}\label{th_mean_rnd_bi}
The mean distribution of fragments in random bicomponent fragmentation is
\begin{empheq}[box=\mybluebox]{equation}
\label{mean_dstr_bi}
    \frac{\ens{n_{a,b}}}{N} = \binom{a+b}{a} 
    \frac{\Omega^{(2)}_{m_A-a,m_B-b;N-1}}{\Omega^{(2)}_{m_A,m_B;N}}
\end{empheq}
\proof The proof follows in the steps of the one-component problem. We express the multiplicity and the partition function in the form
\begin{gather}
   \omega(\mathbf{n}) = 
   N!\prod_{a=0}^\infty\prod_{a=b}^\infty \frac{\alpha_{a,b}^{n_{a,b}}}{n_{a,b}!} , 
   \\
   \Omega^{(2)}_{m_A,m_B;N} 
   = 
   N!\sum_\mathbf{n}\prod_{a=0}^\infty\prod_{a=b}^\infty 
   \frac{\alpha_{a,b}^{n_{a,b}}}{n_{a,b}!}
\end{gather}
With $\alpha_{a,b} = \binom{a+b}{a}$ we recover the result for random fragmentation but for the derivation we treat $\alpha_{a,b}$ as a variable. Following the same procedure that led to Eq.\ (\ref{mean_dstr}) we now obtain
\begin{equation}
\label{mean_dstr_bi_1}
   \frac{\ens{n_{a,b}}}{N} 
   = \alpha_{a,b}\frac{\Omega_{m_A-a,m_B-a;N-1}}{\Omega_{m_A,m_B;N}}. 
\end{equation}
To arrive at this result we note that differentiation of the partition function with respect to $\alpha_{a,b}$ by analogy to Eq.\ (\ref{dlog_omega_dak_2}) amounts to removing one cluster that contains $a$ units of $A$ and $b$ units of $B$, thus producing the partition function $\Omega_{m_A-a, m_B-b; N-1}$ in the numerator of Eq.\ (\ref{mean_dstr_bi_1}). Setting $\alpha_{a,b} = \binom{a+b}{a}$ we obtain Eq.\ (\ref{mean_dstr_bi}). \qed
\end{theorem}

\begin{figure}
\begin{center}
\includegraphics[width=2in]{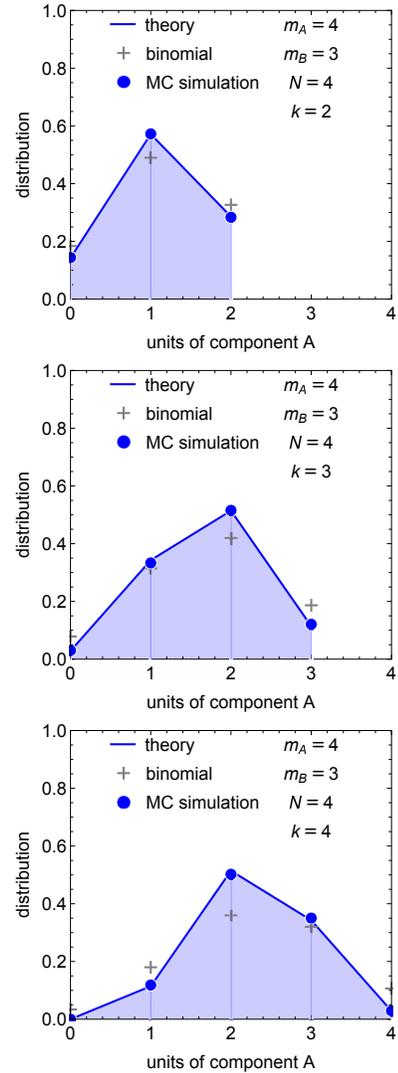}
\end{center}
\caption{The compositional distribution $\ens{c_{a|k}}$ of  bicomponent particle with $m_A=4$, $m_B=3$ into $N=4$ fragments. Distributions are shown for sieve-cut sizes $k=2, 3$ and 4. Lines are from Eq.\ (\ref{mean_dstr_bi}) and points are from MC simulation after $20,000$ fragmentation events. }
\label{fig2}
\end{figure}

\paragraph*{Alternative proof}
The mean distribution can be obtained by a mean-field argument beginning with the observation that the mean color-blind distribution is the same as in the one-component case. This follows from the fact that the choice of the points at which the string o beads is cut is independent of the compositional makeup of the particle (Fig.\ \ref{fig1}).   Thus $\ens{n_k}$ is given by Eq.\ (\ref{mean_dstr}) with $m = m_A+m_B$:
\begin{equation}
\label{mean_size_dstr_bi}
   \frac{\ens{n_k}}{N} 
   = \frac{\Omega^{(1)}_{m_A+m_B-a-b;N-1}}{\Omega^{(1)}_{m_A+m_B;N}}
\end{equation}
We obtain the compositional distribution by the following construction. Imagine that all possible distributions are stacked vertically to form a table so that column 1 contains the first fragment in all distributions, column 2 contains all second fragments and so on. All columns are permutations of each other (this follows from the construction of the fragments illustrated in Fig.\ \ref{fig1}) and since all permutations are equally likely (this follows from the condition of random fragmentation), all columns have the same fragment and compositional distribution, therefore we only need to consider one of them.  The equivalent problem now is this: count the number of ways to select $a$ beads from a pool of $m_A$ $A$'s and $k-a$ beads from a pool of $m_B$ $B$'s and take its ratio over the total number of ways to pick $k$ beads:
\begin{equation}
\label{mean_dstr_compo}
   \ens{c_{a|k}} 
      = \left.
         \binom{m_A}{a}\binom{m_B}{k-a}\right/
         \binom{m_A+m_B}{k} .     
\end{equation}
The mean distribution then is the product of the size and compositional distributions:
\begin{equation}
\label{mean_dstr_size_compo_bi}
   \ens{n_{a|k}} = \ens{n_k} \ens{c_{a|k}} ,
\end{equation}
or
\begin{empheq}[box=\mybluebox]{equation}
\label{mean_dstr_bi_alt}
    \frac{\ens{n_{a,b}}}{N} = 
    \frac{\binom{m_A}{a}\binom{m_B}{b}}{\binom{m_A+m_B}{a+b}}
    \frac{\Omega^{(1)}_{m_A+m_B-a-b;N-1}}{\Omega^{(1)}_{m_A+m_B;N}}
\end{empheq}
It is straightforward algebra to show that this is equivalent to Eq.\ (\ref{mean_dstr_bi}). 
For $m_A,m_B\gg 1$, $a+b\ll m_A+m_B$ the compositional distribution goes over to the binomial:
\begin{equation}
\label{binomial}
    \frac{\binom{m_A}{a}\binom{m_B}{b}}{\binom{m_A+m_B}{a+b}}
   \to
   \binom{k}{a} \phi_A^a \phi_B^b,
\end{equation}
with $\phi_A=m_A/(m_A+m_B)$, $\phi_B=1-\phi_A$. Figure (\ref{fig2}) shows compositional distributions for a bicomponent particle with $m_A=4$ units of $A$ and $m_B=3$ units of $B$. As a means of a demonstration we show the results of a Monte Carlo simulation, which are seen to be in excellent agreement with theory. The binomial distribution, also shown for comparison, is only in qualitative agreement because the fragment masses are small and the conditions for asymptotic behavior are not met in this case. 

\subsection{Any number of components}
Extension to any number of components follows in a straightforward manner from the bicomponent case. Suppose the parent particle consists of $K$ components $A, B\cdots$ and contains $m_A$ units of $A$, m$_B$ units of $B$ and so on.  The distribution of fragments is now expressed by the $K$-dimensional vector $\mathbf{n} = \{n_{a,b\cdots}\}$ that gives the number of fragments that contain $a$ units of component $A$, $b$ units of $B$ etc. This distribution satisfies 
\begin{align}
\label{multi_constraint1}
   & \sum_{a,b\cdots} n_{a,b\cdots} = N \\
\label{multi_constraint2}
   & \sum_{a,b\cdots} z n_{a,b\cdots} = m_Z;\quad z=a,b\cdots
\end{align}
where $m_Z$ is the mass of component $z=a,b\cdots$ in the parent particle. The set all distributions that satisfy the above conditions constitutes the ensemble of all distributions that are produced by the fragmentation of the parent particle into $N$ integer fragments. 

Random fragmentation is once again implemented as shown in Fig.\ \ref{fig1}: Given a string of colored beads we cut it at $N-1$ random points to produce $N$ fragments. All permutations of the beads are equally probable. Accordingly, all ordered lists of fragments (configurations) are equally probable. The number of configurations is
\begin{empheq}[box=\mybluebox]{equation}
\label{Omega_multi}
   \Omega^{(K)}_{\mathbf{m}; N}
   = \mathbf{m!}\, \Omega^{(1)}_{m;N}
   = \left(\frac{m!}{m_A! m_B! \cdots}\right) \binom{m-1}{N-1} , 
\end{empheq}
where $m = m_A+m_B+\cdots$ is the total mass of the particle and $\mathbf{m!}$ is the multinomial coefficient of the compositional vector $\mathbf{m}=(m_A,m_B\cdots)$ of the parent.  The multinomial coefficient of the compositional vector is the number of permutations of the string of beads and the binomial factor is the number of ways to cut it into $N$ pieces. The multiplicity $\omega (\mathbf{n})$ of distribution $\mathbf{n}$ is the number of configurations with that distribution and is given by the number of permutations in the order of fragments and in the order of components within each fragment:
\begin{equation}
\label{omega_multi}
   \omega(\mathbf{n}) 
   = 
   \mathbf{n!} \prod_{a,b\cdots}
   \left(
   \frac{(a+b+\cdots)!}{a! b! \cdots}
   \right)^{n_{a,b\cdots}}
   =
   \mathbf{n!}\prod_\mathbf{c}\left(\mathbf{c!}\right)^{n_\mathbf{c}}
\end{equation}
where $\mathbf{c!}$ is the multinomial factor of the compositional vector $\mathbf{c} = (a,b,c\cdots)$ of the fragment. The probability of distribution $\mathbf{n}$ is its multiplicity normalized by the total number of configurations
\begin{equation}
\label{P_multi}
   P(\mathbf{n}) = 
   \frac{\mathcal \omega(\mathbf{n})}{\Omega^{(K)}_{\mathbf{m}; N}}   
\end{equation}
with
\begin{equation}
\label{norm_multi}
   \Omega^{(K)}_{\mathbf{m}; N} = \sum_\mathbf{n} \omega(\mathbf{n}) .
\end{equation}
The normalizing factor $\Omega^{(K)}_{m_A,m_B\cdots; N}$ is the partition function of the ensemble of distributions that satisfy the constraints in Eqs.\ (\ref{multi_constraint1}) and (\ref{multi_constraint2}). 

The mean distribution of fragments is
\begin{empheq}[box=\mybluebox]{equation}
    \frac{\ens{n_\mathbf{c}}}{N} 
    =
    \mathbf{c}!
    \frac{\Omega^{(K)}_{\mathbf{m}-\mathbf{c};N-1}}{\Omega^{(K)}_{\mathbf{m};N}}
\end{empheq}
and is the generalization of (\ref{mean_dstr_bi}). Alternatively, the mean distribution can be expressed by analogy to Eq.\ (\ref{mean_dstr_size_compo_bi}) as the product of the color blind distribution with a mean compositional distribution:
\begin{empheq}[box=\mybluebox]{equation}
\label{mean_dstr_multi}
     \frac{\ens{n_\mathbf{c}}}{N}
   = \frac{\ens{n_k}}{N}\ens{c_{\mathbf{c}| k}} . 
\end{empheq}
The mean color-blind size distribution $\ens{n_k/N}$ is the same as in one-component fragmentation,
\begin{equation}
\label{mean_size_dstr_multi}
      \frac{\ens{n_k}}{N} = \frac{\Omega^{(1)}_{m-k; N-1}}{\Omega^{(1)}_{m;N}},\quad
      k=a+b+\cdots,
\end{equation}
with $m=m_A+m_B+\cdots$ and $\ens{c_{a,b\cdots|k}}$ is the conditional probability that the compositional vector of fragment size $k$ is $\mathbf{c} = (a,b\cdots)$:
\begin{equation}
c_{\mathbf{c}| k} = 
   \left.
   \binom{m_A}{a}
   \binom{m_B}{b}\cdots
   \right/
   \binom{m}{a+b+\cdots} .
\end{equation}
This is the generalization of Eq.\ (\ref{mean_dstr_compo}). 

\section{Non random fragmentation}
In random fragmentation we produce permutations of the configuration  of fragments at random and accept them with uniform probability 1.  We now bias the acceptance of the permutation by a functional $W(\mathbf{n})$ of the fragment distribution such that the probability of fragment distribution $\mathbf{n}$ in  is
\begin{equation}
\label{P_multi_bias}
   P(\mathbf{n}) =\frac{\omega(\mathbf{n})  W(\mathbf{n})}{\tilde\Omega^{(K)}_{m_A,m_B;N}} 
\end{equation}
with 
\begin{equation}
\label{Omega_multi_bias}
   \tilde\Omega^{(K)}_{m_A,m_B;N} = \sum_\mathbf{n} \omega(\mathbf{n})  W(\mathbf{n})  .
\end{equation}
with $\omega(\mathbf{n})$ from Eq.\ (\ref{omega_multi}).
These are the general forms of Eqs.\ (\ref{P_multi}) and (\ref{norm_multi}) in the bicomponent case and are easily extended to any number of components. Here $\omega(\mathbf{n})$ is the intrinsic multiplicity of $\mathbf{n}$ in the ensemble of fragments, while the product $\omega(\mathbf{n}) W(\mathbf{n})\doteq \tilde\omega(\mathbf{n})$ is its apparent (biased) multiplicity as weighted by the bias functional and distinguished by the tilde. Similarly, the partition function $\tilde\Omega$ is the summation of the apparent (biased) multiplicities of all distributions in $\mathscr E_{m_A,m_B;N}$. With $W=1$ we recover the random case. In this sense ``random'' and ``unbiased'' both refer to the case of uniform bias. 

\subsection{Linear ensemble}

The bias functional $W$ will remain unspecified. This allows us to choose the bias so as to produce any desired distribution of fragments. A special but important case is when $W$ is of the factorizable form
\begin{equation}
\label{W_linear}
   W(\mathbf{n}) 
      = \prod_a \prod_b (w_{a,b})^{n_{a,b}}, 
\end{equation}
where $w_{a,b}$ depend on $a$ and $b$ but not on the fragment distribution itself. The log of the bias is then a linear function of $\mathbf{n}$:
\begin{equation}
   \log W(\mathbf{n}) = \sum_a \sum_b \log w_{a,b} n_{a,b} . 
\end{equation}
The result states that the log of the bias is homogeneous functional of $\mathbf{n}$ with degree 1, i.e. $\log(\lambda \mathbf{n}) = \lambda \log W(\mathbf{n})$ for any $\lambda>0$.  We refer to this case as linear bias with the understanding that linearity actually refers to the log of $W$. 

\begin{theorem}
The mean distribution of fragments under the bias in Eq.\ (\ref{W_linear}) is
\begin{empheq}[box=\mybluebox]{equation}
\label{mean_dstr_multi_bias}
    \frac{\ens{n_{a,b}}}{N} = w_{a,b}\binom{a+b}{a} 
    \frac{\tilde\Omega^{(2)}_{m_A-a,m_B-b;N-1}}{\tilde\Omega^{(2)}_{m_A,m_B;N}}
\end{empheq}
with
\begin{equation}
   \tilde\Omega^{(2)}_{m_A,m_B;N} 
   =
   N!\sum_\mathbf{n} 
   \prod_{a=0}^\infty
   \prod_{b=0}^\infty
   \frac{w_{a,b}^{n_{a,b}}}{n_{a,b}!} \binom{k}{a}^{n_{a,b}}.
\end{equation}

\proof Using  Eq.\ (\ref{W_linear}) the  apparent multiplicity $\tilde\omega(\mathbf{n})$ of distribution $\mathbf{n}$ is
\begin{gather}
   \tilde\omega(\mathbf{n}) = 
   N!\prod_{a=0}^\infty\prod_{a=b}^\infty 
   \frac{(\alpha_{a,b})^{n_{a,b}}}{n_{a,b}!} , 
\end{gather}
with 
\begin{equation}
   \alpha_{a,b} = w_{a,b} \binom{k}{a} . 
\end{equation}
The result then follows directly from Theorem \ref{th_mean_rnd_bi}. \qed

\end{theorem}

\subsection{Composition-independent bias}
If the bias factors are of the form $w_{a,b} =  g(a+b)$, where $g(x)$ is a function of a single variable, the acceptance probability of a configuration of fragments depends on the mass $k=a+b$ of the fragment but not on its composition. This leads to a simple expression for the mean distribution by the following argument. With reference to Fig.\ \ref{fig1}, fix the points where the string is cut; this amounts to fixing the size distribution of the fragments. All permutations of the colored beads are equally probable because they have the same value of $W$. Accordingly, the compositional distribution is the same as in the random case and is given by Eq.\ (\ref{mean_dstr_compo}). The size distribution on the other hand is biased and is the same as when the same bias is applied to one-component distribution. The final result is
\begin{equation}
   \frac{\ens{n_{a,b}}}{N} 
   = 
   \frac{\binom{m_A}{a}\binom{m_B}{k-a}}{\binom{m_A+m_B}{k}}
   \frac{\ens{n_k}}{N} ,
\end{equation}
where $\ens{n_k}$ is the one-component size distribution under bias $w_{a,b}=g(a+b)$,
\begin{equation}
   \frac{\ens{n_k}}{N} 
    =
   g(k) 
   \frac{\tilde\Omega^{(1)}_{m_A+m_B-k; N-1}}{\tilde\Omega^{(1)}_{m_A+m_B; N}}
\end{equation}
with
\begin{equation}
   \tilde\Omega^{(1)}_{m_A+m_B; N}
   =
   N!\sum_\mathbf{n} \prod_{k=1}^\infty \frac{g(k)^{n_k}}{n_k!}
   .
\end{equation}
Except for special forms of $g(k)$ the partition function will not be generally available in closed form. Table \ref{tbl_Omega} summarizes three cases for which exact results are possible. All three cases are associated with distributions encountered in binary aggregation \citep{Matsoukas:springer_2019}. The partition functions in cases 1 and 2 refer to the constant and sum kernels, respectively, and are exact; Case 3 is associated with the product kernel and gives the asymptotic limit of the partition function for $m,N\gg1$, $m/N<2$, conditions that refer to the pre-gel state \citep{Matsoukas:SR15}. 

In the general case $w_{a,b}$ depends on both $a$ and $b$ explicitly and affects both the size and compositional distributions. This case will be demonstrated by simulation in the next section.

\begin{table}
\caption{Closed form results for three composition-independent bias functionals}
\label{tbl_Omega}
\begin{equation*}
\renewcommand{\arraystretch}{3}
\begin{array}{l *2 {>{\displaystyle}c} }
 &  \makebox[80pt][c]{$ w_{a,k-a}$} & \Omega^{(1)}_{m;N}  \\
\hline
   \text{Case 1} &
   1 & 
   \binom{m-1}{N-1}
\\
   \text{Case 2} &
   \frac{2(2k)^{k-2}}{k!} & 
   m^{m-N}\frac{N!}{m!}\binom{m-1}{N-1} 
\\
   \text{Case 3}^\ddagger &
   \frac{2(2k)^{k-2}}{k!} & 
   \left(m^{m-N}\frac{N!}{m!}\right)^2\binom{m-1}{N-1} 
\\
\hline
\multicolumn{3}{l}{^\ddagger\text{asymptotically for $m, N\gg 1$, $m/N<2$}}
\end{array}    
\end{equation*}
\end{table}

\section{Simulation of biased fragmentation}
Except for certain special forms of the bias the mean fragment distribution cannot be calculated analytically and the only recourse is stochastic simulation. Here we describe a Monte Carlo (MC) algorithm for sampling the ensemble of distributions. We will then use this method to demonstrate result for two cases of biased bicomponent fragmentation. 

\subsection{Monte Carlo sampling by exchange reaction}

Suppose $\mathcal C = ((a_1,b_1),\cdots (a_N,b_N))$ is a configuration of $N$ fragments such that fragment $i$ contains $a_i$ units of component $A$ and $b_i$ units of component $B$. Since all configurations with the same distribution have the same probability, given in Eq.\ (\ref{P_multi_bias}), the probability of configuration is
\begin{equation}
   P(\mathcal C) = \frac{W(\mathcal C)}{\Omega^{(2)}_{m_A,m_B;N}}
\end{equation}
where $W(\mathcal C)=W(\mathbf{n})$ and $\mathbf{n}$ is the distribution of the configuration. If the bias functional is of the linear form in Eq.\ (\ref{W_linear}), its value on $\mathcal C$ is
\begin{equation}
   W(\mathcal C) = \prod_{i=1}^{N} w_{a_i,b_i} . 
\end{equation}
Suppose that two fragments $i$ and $j$ exchange mass according to the reaction
\begin{equation}
\label{rxn}
   (a_i,b_i) + (a_j,b_j) \to (a'_i,b'_i) + (a'_j,b'_j) 
\end{equation}
under the conservation conditions $a_i+a_j=a'_i+a'_j$ and $b_i+b_j=b'_i+b'_j$. This amounts to a transition between configurations, $\mathcal C \to \mathcal C'$, with equilibrium constant
\begin{equation}
   \mathcal K_{\mathcal C\to\mathcal C'} 
   = \frac{P(\mathcal C')}{P(\mathcal C)}
   = \frac{W(\mathcal C')}{W(\mathcal C)}
   =\frac{w_{a'_i,b'_i} w_{a'_j,b'_j}}{w_{a_i,b_i} w_{a_j,b_j}} .
\end{equation}
This has the form of a chemical equilibrium constant for the exchange reaction in (\ref{rxn}) with $w$ playing the role of the activity of the species. The ensemble of fragment distributions may then be sampled via Monte Carlo by direct analogy to chemical reactions \citep{Matsoukas:E19}. 
Given a configuration of fragments, pick two at random and implement an exchange of mass by a random amount. The resulting configuration is accepted by the Metropolis criterion: accept if $\texttt{rnd}\leq K_{\mathcal C\to\mathcal C'}$, where $\texttt{rnd}$ is a random number uniformly distributed in $(0,1)$; otherwise reject. With $W=1$ every exchange reaction is accepted, which amounts to random fragmentation. 
We implement the random exchange between fragments as follows. We represent fragments as a list of 1's (representing component $A$) and $0$'s (component $B$). We pick two clusters $i$ and $j$ and random. We merge them into a single list, randomize the order of components, and break into two new fragments by picking a break point at random. The randomization of the order of components in the merged list ensures that all permutations are considered with equal probability.

\begin{figure}
\begin{center}
\includegraphics[width=3.0in]{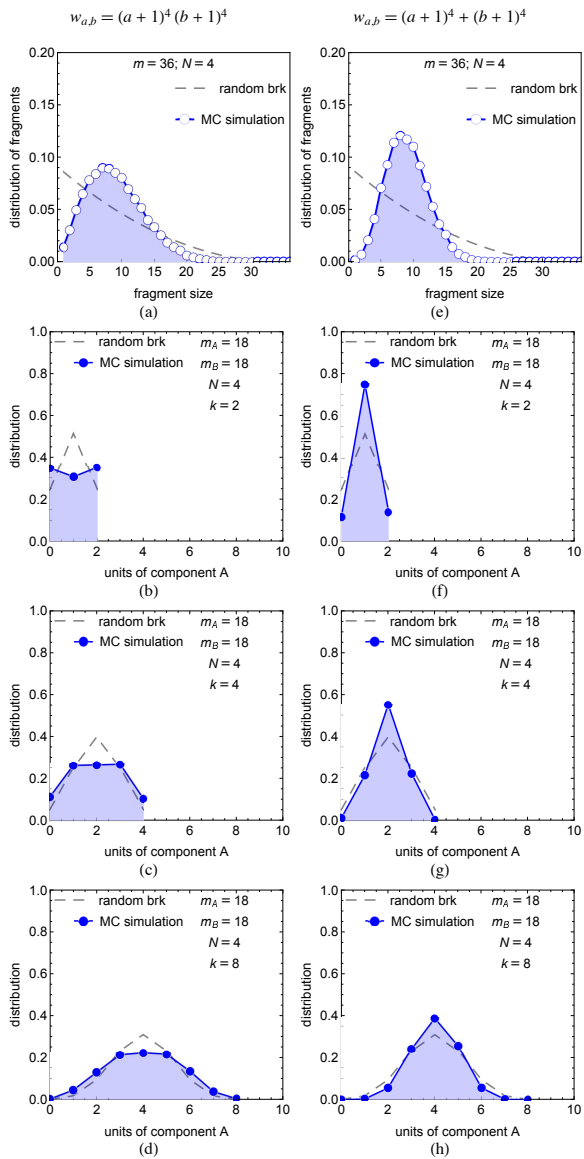}
\end{center}
\caption{Size and compositional distributions at fragment sizes $k=2,4$ and 8 for two bias functionals: (a)--(d): $w_{a,b} = (1+a)^4(1+b)^4$; (e)--(h): $w_{a,b} = (1+a)^4+(1+b)^4$. In both cases the particle contains $m_A=18$ units of $A$, $m_B=18$ units of $B$ and breaks into $N=4$ fragments. }
\label{fig3}
\end{figure}

\subsection {Two examples}
In random fragmentation ($w_{a,b}=1$) the compositional distribution is given by Eq.\ (\ref{mean_dstr_compo}). We may choose the bias functional so as to produce deviations in either direction relative to the random case. It is possible to produce positive deviations (preferential segregation of components in the fragments relative to random mixing) or negative deviations (more intimate mixing than in random mixing). We demonstrate both behaviors using the two examples below:
\begin{enumerate}
\item Case I (positive deviations)
\begin{equation}
   w_{a,b} = (a+1)^\alpha + (b+1)^\alpha
\end{equation}
\item Case II (negative deviations)
\begin{equation}
   w_{a,b} = (a+1)^\alpha(b+1)^\alpha
\end{equation}
\end{enumerate}

In Case I the fragment bias $w_{a,b}$ is an additive function of the amounts of the two components. Considering that $a+b$ is constrained by mass balance, the fragment bias is large for fragments that are rich in either component but small for fragments that are relatively mixed. This ought to favor the formation of fragments in which the components are relative segregated. The fragment bias in Case II is a multiplicative function of the amounts of the two components. It is large in fragments that contain both components but quite small if one component is present in excess of the other. This form ought to produce fragments that are better mixed than fragments produced by random fragmentation. 

We test these behaviors in Fig.\ \ref{fig3} which shows results for $\alpha = 4$. In this example the particle contains an equal number of units of each component, $m_A=m_B=18$, and breaks into $N=4$ pieces.  
In both cases the size distribution deviates  from that in random fragmentation. 
Indeed the bias influences both the size and the compositions distribution unless its is of the special form $w_{a,b} = g(a+b)$. 
Compositional distributions are shown for sieve-cut masses $k=2$, 4 and 8. The additive bias (Case I) produces distributions that are more spread out relative to the random case. For $k=2$, in particular, the compositional distribution is inverted relative to the random case, indicating strong segregation as the majority of fragments contains pure component $A$ or $B$ and only few fragments in this size contain both components.  As the fragment size increases the separation of components is less strong but always present, as indicated by the fact that the random distribution is always narrower. The opposite behavior is observed in Case II: distributions are narrower than those in random fragmentation, especially at the smaller fragment sizes. 

As a general trend in both cases, small fragments are less mixed while large fragments progressively approach the distribution of random fragmentation. This is because there are not enough units of each component to produce large fragments that consist predominantly of one component. This limitation is not present when the fragment size is small.

\section {Discussion and Conclusions}
We have presented a treatment of multicomponent fragmentation on the basis of random fragmentation in combination with an appropriate functional that biases the ensemble of feasible distributions. The two key notions in this treatment are the set of feasible distributions and the multiplicity of distribution within this set as established by the rules that define ``random'' fragmentation. In the random-fragmentation ensemble  distributions are proportional to their multiplicity. This problem is analytically tractable and we have presented its solution for any number of component and number of fragments. A third key notion is that of the  bias functional that modulates the probability of distributions of feasible distributions and allows us to obtain fragment distributions other than that of random fragmentation. 

It should be pointed out that the random case is not endowed with universal physical significance but applies in certain cases such as the linear chain in Fig.\ \ref{fig1}. In this particular case selecting the bonds to break at random might be a reasonable physical model, as \citet{Montroll:JCP40} explain. The primary significance of random fragmentation is mathematical. Similar to the ``fair coin'' or the ``ideal solution,'' it provides an analytically solvable baseline (``reference state'') from which to calculate deviations. The mathematical tool that quantifies these deviations is the bias functional. This functional permits the systematic construction of distributions that exhibit any degree of deviation from the random case. 
This is main result of this formulation. The fragment distribution per fragmentation event is one of two elements required in order to build a population balance model of a fragmentation process. The other element is the rate at which particles break up. This question is not addressed here beyond the generic observation that this rate must be a function of the compositional vector $\mathbf{m} =(m_A,m_B\cdots)$ of the particle. 

In single-component fragmentation the quantity of interest is the mean size distribution of the fragments. In multicomponent systems we are additionally concerned with the compositional distribution. This introduces a new dimension to the problem and raises questions of mixing and unmixing of components. Do fragments inherit the compositional characteristics of the parent particle? Do they become progressively more well mixed or less?
Both behaviors are possible and are quantified via the bias functional $W$. This functional is where the mathematical theory of fragmentation presented here makes contact with the physical mechanisms that lead to the disintegration of material particles. To make this connection quantitatively, one must begin with the a physical model of fragmentation that assigns probabilities to all possible distributions of fragments that can be generated. This is a major undertaking and is specific to the particular problem that is being considered. The point we wish to make is that the formulation presented here offers an entry point to physics via the bias functional. 

Lastly, the connection to statistical mechanics should not be lost. We have constructed an ensemble whose fundamental element (``microstate'') is a the ordered configuration of fragments; its total number in the ensemble is the partition function. The higher-level stochastic variable (the observable) is the distribution of fragments and its probability is determined by its multiplicity in the ensemble. The form of the probability in Eq.\ (\ref{prob_b2}), also known as Gibbs distribution \citep{Berestycki:07},  is encountered in time reversible processes as well as in population balances of aggregation and breakup \citep{Hendriks:ZPCM85B,Durrett:JTP99,Berestycki:EJP04,Berestycki:07,Kelly:2011}.  The derivation of the mean distribution in the random case follows in the steps of the Darwin-Fowler method \citep{Schrodinger:89}.  
Additionally, the compositional distribution in random breakup is given asymptotically by the binomial distribution in Eq.\ (\ref{binomial}). This establishes a reference for compositional interactions analogous to that of the ideal solution in thermodynamics. In fact, the Shannon entropy of the binomial distribution is the ideal entropy of mixing when two pure components coalesce into a single particle that contains mass fraction $\phi_A$ of component $A$.  These connections are not coincidental. Biased sampling from a distribution generates a probability space of distributions and when the base distribution is exponential, this ensemble obeys thermodynamics \citep{Matsoukas:E19}. In fragmentation the base distribution is a multicomponent exponential: the size distribution in Eq.\ (\ref{mean_dstr}) goes over to the exponential distribution when $m,N\gg 1$. In this limit the ensemble of fragments becomes mathematically equivalent to a thermodynamic ensemble of two components with interactions that lead to positive or negative deviations relative to ideal solution. 

\bibliography{tm,statMech,pbe}
\bibliographystyle{apsrev}

\end{document}